\documentclass[twoside]{llncs}
\usepackage{fancyhdr}           
\usepackage{ifthen}
\fancypagestyle{plain}{%
\fancyhf{}               

}
\newcommand\settitle[2][]{%
 \title{#2}
 \ifthenelse{\equal{#1}{}}%
  {\fancyhead[RO]{\nouppercase #2 \qquad \thepage}}%
  {\fancyhead[RO]{\nouppercase #1 \qquad \thepage}}%
}
\newcommand\setauthors[2]{%
 \author{#2}
  {\fancyhead[LE]{\thepage \qquad \nouppercase #1}}%
}
\newcommand{\keywords}[1]{\noindent {\bf Keywords:}\ #1}

\begin{document}
\settitle[Semantic Integration in the IFF]
         {Semantic Integration \\ in the \\ Information Flow Framework}
\setauthors{R.E. Kent}
           {Robert E. Kent}
\institute{Ontologos\\
  550 SW Staley Dr, Pullman WA 99163, USA\\
\email{rekent@ontologos.org}}
\date{}
\maketitle
\thispagestyle{plain}


\begin{abstract}
The Information Flow Framework (IFF) \cite{iff:homepage}
is a descriptive category metatheory currently under development,
which is being offered as the structural aspect of the Standard Upper Ontology (SUO). 
The architecture of the IFF is composed of metalevels, namespaces and meta-ontologies.
The main application of the IFF is institutional: 
the notion of institutions and their morphisms are being axiomatized in the upper metalevels of the IFF, 
and the lower metalevel of the IFF has axiomatized various institutions
in which semantic integration has a natural expression as the colimit of theories.
Some of the ideas used in this paper
first appeared in papers by Joseph Goguen \cite{goguen:05} and the author \cite{kent:iswc03},
and discussions on the SUO email list.
See also the companion paper \cite{goguen:dagstuhl}.
\end{abstract}

\keywords{descriptive category metatheory, institutions, semantic integration}

\begin{quotation}
``\emph{Philosophy cannot become scientifically healthy without an immense technical vocabulary. 
We can hardly imagine our great-grandsons turning over the leaves of this dictionary without amusement over the paucity of words with which their grandsires attempted to handle metaphysics and logic. Long before that day, 
it will have become indispensably requisite, too, 
that each of these terms should be confined to a single meaning which, however broad, must be free from all vagueness. 
This will involve a revolution in terminology; 
for in its present condition a philosophical thought of any precision can seldom be expressed without lengthy explanations.}'' 

\noindent --- Charles Sanders Peirce, Collected Papers 8:169
\end{quotation}

\section{The Information Flow Framework}

The IEEE P1600.1 Standard Upper Ontology (SUO) project aims to specify an upper ontology that will provide a structure and a set of general concepts upon which object-level ontologies can be constructed.
The Information Flow Framework (IFF) \cite{iff:homepage} is a descriptive category metatheory that represents the structural aspect of the SUO containing meta, generic and abstract concepts.
To be independent of the particular logic used in object-level ontologies,
the IFF represents and manipulates ontological structures within the metatheory of institutions.
The IFF has work-in-progress axiomatizations for (amongst others) 
the institutions of information flow ({\bfseries IF}), equational logic ({\bfseries EQL}), order-sorted first order logic ({\bfseries FOL}) and the common logic standard ({\bfseries CL}),
and is developing an axiomatization for the metatheory of institutions itself.
Its institutional approach to logical semantics provides a principled framework for the modular design of object-level ontologies;
in particular, for the ``lattice of theories'' approach to ontological organization.
A major requirement of the SUO, 
called the ``lattice of theories'', 
is to create a framework which can support an open-ended number of theories (potentially infinite) organized in a lattice together with systematic metalevel techniques for moving from one to another, for testing their adequacy for any given problem, and for mixing, matching, combining, and transforming them to whatever form is appropriate for whatever problem anyone is trying to solve (John Sowa \cite{sowa:00} and the SUO archive).
Within the IFF representation, 
the lattice of theories is the fibring or indexing of the context of theories by the context of signatures (aka languages)
--- all definable within institutions. 
Semantic integration of ontologies can be represented by 
(1) aligning ontologies within a diagram of theories and 
(2) fusing aligned ontologies via the colimit of that diagram of theories.

The IFF has gone through two phases of development, and is now well within its third phase,
which involves reorganization of the IFF core hierarchy using the ``adjunctive axiomatization'' technique as illustrated by Lawvere and Rosebrugh \cite{lawvere:rosebrugh:03}.
A fourth phase is envisioned in the future, where the concepts of fibrations and indexed categories will be axiomatized.
These are important for the axiomatizations of institutions and fibring logics.
During the IFF development,
four concepts have eventually emerged as important.
In chronological order these are
(1) the conceptual warrant principle,
(2) the categorical design principle,
(3) the institutional logic principle, and
(4) the metastack concept.
\begin{itemize}
\item 
\noindent {\bfseries Conceptual Warrant:} 
\emph{All IFF terminology should require conceptual warrant for their existence:
any term that appears in (and is axiomatized by) a metalanguage should reference a concept needed in a lower metalevel or object level axiomatization.}
\newline The principle of conceptual warrant originated in phase one of the IFF development.
Warrant means evidence for or token of authorization. 
Conceptual warrant is an adaptation of the librarianship notion of literary warrant.
According to the Library of Congress, its collections serve as the literary warrant 
(i.e., the literature on which the controlled vocabulary is based) 
for the Library of Congress subject headings system.
In the same fashion,
the object-level and lower metalevel terminology of the IFF serves as the conceptual warrant 
for the IFF upper metalevel axiomatization. \\
\item 
\noindent {\bfseries Categorical Design:}
\emph{The design of a module should adhere to the property that its axiomatic representation is strictly category-theoretic:
All axioms use terms from the metalanguage at that metalevel.
No axioms use explicit logical notation:
no variables, quantifications or logical connectives are used.}
\newline The principle of categorical design originated in phase two of the IFF development.
The goal of this principle is to simplify the IFF axiomatization 
--- first order expression would be reduced to term-rewriting.
The peripheral (non-core) modules in the lower IFF metalevel have the tripartite form:
outer category namespace, inner object and morphism namespaces. 
The outer namespace fully conforms to the categorical design principle.
The inner namespaces conform to it to a great extent (80--90\%). 
\\
\item 
\noindent {\bfseries Institutional Logic:}
\emph{All logics used in the IFF application should be formulated as institutions.}
\newline The principle of institutional logic\footnote{Suggested to the author by Joseph Goguen (personal communication).} originated in phase three of the IFF development.
The theories of information flow and formal concept analysis
(and hence, effectively the theory of institutions) 
have been used throughout the IFF development.
This use has centered on the ``truth construction'', which is institutional (see Subsection~\ref{subsec-truth-construction}).
As noted in \cite{goguen:logic},
truth is not dyadic between models and sentences,
but triadic between models, sentences and signatures.
This corresponds to the contextual dependency in the semiotics of Charles Sanders Peirce,
which uses signs, objects and interpretants.
In formal concept analysis,
triadic concept lattices \cite{lehmann:wille} have been used to formalize this.
However,
the representation of contexts as the logics of institutions has greater advantages.
\\
\item 
\noindent {\bfseries Metastack:}
\newline The idea of the IFF metastack also originated in phase three of the IFF development.
The IFF metastack is the IFF core hierarchy.
The content of the IFF metastack is the axiomatization for sets, functions and binary relations 
in four core modules at four different set-theoretic levels, 
the ``small'' in the lower metalevel, the ``large'' in the upper metalevel, 
the ``very large'' in the top metalevel and the ``generic'' in the ur metalevel. 
This includes
an axiomatization for generic categories in the ur metalevel, 
an axiomatization for very large finite limits in the top metalevel, 
an axiomatization for large exponents and finite colimits in the upper metalevel, and
an axiomatization for small subobject classifiers and general limits/colimits in the lower metalevel. 
Although, this effectively distributes topos axioms over the four metalevels,
there is still an effort to follow conceptual warrant.
\end{itemize}

The IFF architecture consists of metalevels, namespaces and meta-ontologies.
Within each level, the terminology is partitioned into namespaces. 
The number of namespaces and the content may vary over time: 
new namespaces may be created or old namespaces may be deprecated, 
and new terminology and axiomatization within any particular namespace may change (new versions).
In addition, within each level, various namespaces are collected together into meaningful composites called meta-ontologies. 
At any particular metalevel, 
these meta-ontologies cover all the namespaces at that level, but they may overlap. 
The number of meta-ontologies and the content of any meta-ontology may vary over time: 
new meta-ontologies may be created or old meta-ontologies may be deprecated, 
and new namespaces within any particular meta-ontology may change (new versions).

The IFF terminology is managed in terms of namespace prefixes
--- each namespace is given a unique prefix (with perhaps a few synonyms) in order to avoid clash of terminology.
The architecture of the IFF namespace mechanism is flat
--- namespace prefixes are like tags:
by using namespace prefixes the complete IFF terminology is the disjoint union of the terminology in the IFF namespaces.
The IFF architecture can be thought of as a two dimensional structure consisting of metalevels,
which are partitioned into top-level namespaces representing basic concepts such as ``category'', ``graph'' or ``institution''.
The various levels are indexed by the natural numbers or their language correlates, starting with the object level indexed by zero.
Overall, various namespaces may have the same name, since they represent the same basic concept at different metalevels.
To locate any namespace one can use its level-concept pair.
For example, the namespace that axiomatizes large categories is located at the ``second'' metalevel and represents the ``category'' concept. 
It is assumed that each basic concept has a particular metalevel that is in common use.
For such namespaces, the level notation need not be used. 

There are thousands of terms in the IFF.
Terms divide into two classes, which we can call ``usable terms'' and ``supporting terms''.
An IFF term, which is defined in a particular namespace on a particular metalevel, 
is a \emph{usable} IFF term when it is used by at least one other term in another namespace on that metalevel or on a level below that one.
An IFF term, which is defined in a particular namespace on a particular metalevel, 
is a \emph{supporting} IFF term when it is used by another term in that same namespace.
Because of conceptual warrant,
all IFF terms should be usable or supporting, and perhaps both.
Hence, all IFF terms are necessary, but most IFF terms are ``conceptually derived''.
This means that they are a conceptual composite of two or more basic IFF terms.
An IFF term is a \emph{basic} IFF term when it is not the conceptual composite of two or more other IFF terms.
Currently, there are about one hundred basic IFF terms.

There are four IFF metalevels: lower, upper, top and ur.
Each metalevel services the level below: 
the ur metalevel services the top metalevel, 
the ur and top metalevels service the upper metalevel, 
the ur, top and upper metalevels service the lower metalevel, and 
the ur, top, upper and lower metalevels service the object-level.
There is one metalanguage associated with each metalevel.
That is, each metalevel has an associated metalanguage, 
whose {\em old} terminology is the terminology of the metalanguage associated with the metalevel immediately above and whose {\em new} terminology is defined by the various meta-ontologies at that metalevel. 
Any metalanguage can be used by the meta-ontologies and ontologies at all lower levels.
There results a hierarchy of IFF metalanguages 
$\mathtt{meta\mbox{-}lower} \sqsupseteq \mathtt{meta\mbox{-}upper} \sqsupseteq \mathtt{meta\mbox{-}top} \sqsupseteq \mathtt{meta\mbox{-}ur} \sqsupseteq \mathtt{metashell}$
coordinated with the IFF metastack, where the logical shell called {\ttfamily metashell} enables a lisp-like first-order expression using connectives and quantifiers with a restricted quantification format.
The {\ttfamily meta-ur} metalanguage is special --- it axiomatizes the IFF metastack.

\section{Semantic Integration with Institutions}

Abstract semantic integration can be defined and axiomatized in the metatheory of institutions,
where it has both an intrinsic and an extrinsic formulation.
In this paper, we discuss the intrinsic formulation in terms of the colimits of theories.
A later paper will discuss the extrinsic formulation.

\subsection{Institutions}

An institution $\Im = \langle \mathsf{Sign}_{\Im}, \mathsf{mod}_{\Im}, \mathsf{sen}_{\Im}, \models_{\Im} \rangle$
\cite{goguen:burstall:92} has
an abstract category $\mathsf{Sign}_{\Im}$ of signatures $\Sigma$,
a sentence fiber functor $\mathsf{sen}_{\Im} : \mathsf{Sign}_{\Im} \rightarrow \mathsf{Set}$
indexing abstract sentences $\mathsf{sen}_{\Im}(\Sigma)$ by signatures $\Sigma$,
a model fiber (reduct) functor $\mathsf{mod}_{\Im} : \mathsf{Sign}_{\Im} \rightarrow \mathsf{CAT}^{\mathsf{op}}$
indexing abstract models $\mathsf{mod}_{\Im}(\Sigma)$ by signatures $\Sigma$, and
a function $\models_{\Im} : |\mathsf{Sign}_{\Im}| \rightarrow \mathsf{REL}$
indexing abstract satisfaction relations
$\models_{\Im, \Sigma} \; \subseteq |\mathsf{mod}|_{\Im}(\Sigma) \times \mathsf{sen}_{\Im}(\Sigma)$ by signatures $\Sigma$.
An institution must satisfy the \emph{satisfaction condition}:
$|\mathsf{mod}|_{\Im}(\sigma)(m_{2}) \models_{\Im, \Sigma_{1}} s_{1}$ 
\underline{iff} 
$m_{2} \models_{\Im, \Sigma_{2}} \mathsf{sen}_{\Im}(\sigma)(s_{1})$,
for any signature morphism $\sigma : \Sigma_{1} \rightarrow \Sigma_{2}$, 
any target model $m_{2} \in |\mathsf{mod}|_{\Im}(\Sigma_{2})$ and any source sentence $s_{1} \in \mathsf{sen}_{\Im}(\Sigma_{1})$.
The satisfaction condition expresses the invariance of truth under change of notation. 
Satisfaction does not use model morphisms, morphisms in $\mathsf{mod}_{\Im}(\Sigma)$,
and hence is expressed in terms of the underlying model functor
$|\mathsf{mod}|_{\Im}
= \mathsf{mod}_{\Im} \circ {\scriptstyle |{-}|}^{\mathsf{op}} : \mathsf{Sign}_{\Im} \rightarrow \mathsf{SET}^{\mathsf{op}}$.
Examples of institutions include:
first order logic with first order structures as models,
many sorted equational logic with abstract algebras as models,
Horn clause logic, and variants of higher order and of modal logic.

For every signature $\Sigma$ of $\Im$,
there is a complete preorder $\mathsf{th}_{\Im}^{\vdash}(\Sigma)$,
whose elements, called $\Sigma$-theories of $\Im$, are subsets of sentences,
whose order is entailment order,
whose join operator is intersection,
and whose meet operator is union.
Elements of theories are called axioms.
We can extend satisfaction to theories:
a $\Sigma$-model $m$ satisfies (is a model of) a $\Sigma$-theory $T$, denoted $m \models_{\Im, \Sigma} T$,
when $m$ satisfies each axiom $s \in T$;
that is, when $m \models_{\Im, \Sigma} s$ for all axioms $s \in T$. 
A theory $T \in \mathsf{th}_{\Im}^{\vdash}(\Sigma)$ entails a sentence $s \in \mathsf{sen}_{\Im}(\Sigma)$,
denoted by $T \vdash_{\Im, \Sigma} s$,
when $m \models_{\Im, \Sigma} T$ implies $m \models_{\Im, \Sigma} s$ for any model $m$.
Such a sentence is called a theorem of $T$.
The set of all theorems of $T$, called the closure of $T$, is denoted by $\mathsf{clo}_{\Im, \Sigma}(T) = T^{\bullet}$.
Closure is monotonic, increasing and idempotent.
If $T_{1} \subseteq T_{2}$ then $T_{1}^{\bullet} \subseteq T_{2}^{\bullet}$. 
Any axiom is a theorem: $T \subseteq T^{\bullet}$.
The closure of the closure is the closure: $T^{\bullet\bullet} = T^{\bullet}$.
A $\Sigma$-theory $T_{1}$ entails a $\Sigma$-theory $T_{2}$ when $T_{1}$ entails every axiom of $T_{2}$;
that is, when $T_{1}^{\bullet} \supseteq T_{2}$.
Then we also say that $T_{2}$ is a generalization of $T_{1}$ or that $T_{1}$ is a specialization of a $T_{2}$,
and denote this entailment order by $T_{1} \vdash_{\Im, \Sigma} T_{2}$.
Entailment is equivalent to closure-subset:
$T_{1} \vdash_{\Im, \Sigma} T_{2}$ iff $T_{1}^{\bullet} \supseteq T_{2}^{\bullet}$.
Hence,
two $\Sigma$-theories are entailment equivalent $T_{1} \equiv T_{2}$ iff
they have the same closure $T_{1}^{\bullet} = T_{2}^{\bullet}$.
In particular,
any theory $T$ is equivalent to its closure in entailment preorder:
$T \equiv T^{\bullet}$.
A $\Sigma$-theory is closed when it is equal to, not just equivalent to, its closure: $T = T^{\bullet}$.
Closed theories $T^{\bullet}$ are in one-one correspondence with equivalence classes $[T]$;
that is, closed theories form a set of representative elements for the entailment equivalence relation.
Let $\mathsf{th}_{\Im}^{\bullet}(\Sigma)$ denote the complete lattice,
whose elements are the closed $\Sigma$-theories of $\Im$, 
whose order is reverse subset inclusion,
whose join operator is intersection,
and whose meet operator is union followed by closure.

Every model $m \in |\mathsf{mod}|_{\Im}(\Sigma)$ has an associated (closed) theory
$\iota_{\Im}^{\bullet}(\Sigma)(m) = \{ s \in \mathsf{sen}_{\Im}(\Sigma) \,|\, m \models_{\Im, \Sigma} s \}$
which is maximal in subset order (minimal in entailment order) for all theories that $m$ satisfies.
This definition gives a model closed embedding function
$\iota_{\Im}^{\bullet}(\Sigma) : |\mathsf{mod}|_{\Im}(\Sigma) \rightarrow \mathsf{th}_{\Im}^{\bullet}(\Sigma)$
and a model entailment embedding function
$\mathsf{mod\mbox{-}th}_{\Im}(\Sigma)= \iota_{\Im}^{\bullet}(\Sigma) \cdot \mathsf{incl}_{\Im, \Sigma} : |\mathsf{mod}|_{\Im}(\Sigma) \rightarrow \mathsf{th}_{\Im}^{\vdash}(\Sigma)$.
Every sentence $s \in \mathsf{sen}_{\Im}(\Sigma)$ has an associated (closed) theory
$\mathsf{sen\mbox{-}th}_{\Im}(\Sigma)(s) = \{ s^{\prime} \in \mathsf{sen}_{\Im}(\Sigma) \,|\, \{s\} \vdash_{\Im, \Sigma} s^{\prime} \}$.
This definition gives a sentence entailment embedding function
$\mathsf{sen\mbox{-}th}_{\Im}(\Sigma) : \mathsf{sen}_{\Im}(\Sigma) \rightarrow \mathsf{th}_{\Im}^{\vdash}(\Sigma)$
and a sentence closed embedding function
$\tau_{\Im}^{\bullet}(\Sigma)
= \mathsf{sen\mbox{-}th}_{\Im}(\Sigma) \cdot \mathsf{clo}_{\Im, \Sigma} : \mathsf{sen}_{\Im}(\Sigma) \rightarrow \mathsf{th}_{\Im}^{\bullet}(\Sigma)$,
since closure is idempotent.

\begin{figure}
\begin{center}
\begin{tabular}{c}
\setlength{\unitlength}{1.0pt}
\begin{picture}(300,180)(0,-20)
\put(0,0){\begin{picture}(240,150)(0,0)
\linethickness{0.5pt}
\put(-50,-37.5){\makebox(100,75){$\mathsf{CAT}^{\mathsf{op}}$}}
\put(50,-37.5){\makebox(100,75){$\mathsf{SET}^{\mathsf{op}}$}}
\put(50,37.5){\makebox(100,75){$\mathsf{CLSN}$}}
\put(125,37.5){\makebox(100,75){$\mathsf{CLG}$}}
\put(190,37.5){\makebox(100,75){$\mathsf{CPOG}$}}
\put(50,112.5){\makebox(100,75){$\mathsf{Set}$}}
\put(125,112.5){\makebox(100,75){$\mathsf{Ord}$}}
\put(125,-37.5){\makebox(100,75){$\mathsf{ORD}^{\mathsf{op}}$}}
\put(190,112.5){\makebox(100,75){$\mathsf{Pre}$}}
\put(-50,37.5){\makebox(100,75){$\mathsf{Sign}_{\Im}$}}
\put(87.5,37.5){\makebox(100,75){$\equiv$}}
\put(0,45){\makebox(100,75){$\mathsf{clsn}_{\Im}$}}
\put(28,18.5){\makebox(100,75){$\mathsf{clg}_{\Im}$}}
\put(29,-13){\makebox(100,75){$\mathsf{cpog}_{\Im}$}}
\put(-15,118){\makebox(100,75){$\exists_{\Im}$}}
\put(-10,82){\makebox(100,75){$\mathsf{sen}_{\Im}$}}
\put(-65,0){\makebox(100,75){$\mathsf{mod}_{\Im}$}}
\put(-19,-10){\makebox(100,75){$|\mathsf{mod}|_{\Im}$}}
\put(60,75){\makebox(100,75){$\mathsf{typ}$}}
\put(88,75){\makebox(100,75){$\mathsf{typ}^{\bullet}$}}
\put(60,10){\makebox(100,50){$\mathsf{inst}$}}
\put(88,10){\makebox(100,50){$\mathsf{inst}^{\bullet}$}}
\put(204,75){\makebox(100,75){$\mathsf{right}^{\vdash}$}}
\put(0,-47.5){\makebox(100,75){$|{-}|^{\mathsf{op}}$}}
\put(87.5,55){\makebox(100,75){$\mathsf{clg}$}}
\put(87.5,23.5){\makebox(100,75){$\mathsf{clsn}$}}
\put(157.5,47.5){\makebox(100,75){$\mathsf{incl}$}}
\put(37,3){\makebox(100,75){$\mathsf{clo}_{\Im}$}}
\put(25,4){\makebox(100,75){\large{$\Uparrow$}}}
\put(108,77){\makebox(100,75){\large{$\Rightarrow$}}}
\put(108,-12){\makebox(100,75){\large{$\Rightarrow$}}}
\put(108,87){\makebox(100,75){$\tau^{\bullet}$}}
\put(108,-2){\makebox(100,75){$\iota^{\bullet}$}}
\put(140,75){\makebox(100,75){$\mathsf{right}^{\bullet}$}}
\put(138,0){\makebox(100,75){$\mathsf{left}^{\bullet}$}}
\put(105,-47.5){\makebox(75,75){\footnotesize{$|{-}|^{\mathsf{op}}$}}}
\put(87.5,120){\makebox(100,75){$|{-}|$}}
\put(175,90){\vector(0,1){45}}
\put(175,62){\vector(0,-1){49}}
\put(155,150){\vector(-1,0){35}}
\put(155,0){\vector(-1,0){35}}
\put(165,87){\vector(-1,1){53}}
\put(165,63){\vector(-1,-1){53}}
\put(20,0){\vector(1,0){60}}
\put(20,75){\vector(1,0){60}}
\put(0,60){\vector(0,-1){45}}
\put(20,90){\vector(4,3){60}}
\put(12,60){\vector(4,-3){70}}
\put(100,60){\vector(0,-1){45}}
\put(100,90){\vector(0,1){45}}
\put(240,90){\vector(0,1){45}}
\put(120,83){\vector(1,0){35}}
\put(155,67){\vector(-1,0){35}}
\put(192,75){\vector(1,0){30}}
\qbezier(2,95)(40,200)(225,152)
\put(225.5,152){\vector(4,-1){0}}
\qbezier(18,67)(100,30)(156,60.5)
\put(156.5,61){\vector(2,1){0}}
\qbezier(15,63)(167.5,-25)(228,60.5)
\put(228.5,61){\vector(3,4){0}}
\end{picture}}
\put(300,0){\begin{picture}(1,150)(0,0)
\put(0,60){\vector(0,-1){45}}
\put(-50,25){\makebox(100,100){$\mathsf{Theory}_{\Im}^{\exists}$}}
\put(-50,-50){\makebox(100,100){$\mathsf{Sign}_{\Im}$}}
\put(-32,-12.5){\makebox(100,100){$\mathsf{base}_{\Im}^{\exists}$}}
\end{picture}}
\end{picture}
\\
\begin{tabular}{c@{\hspace{15pt}}c}
$\begin{array}[t]{|@{\hspace{4pt}}rcl@{\hspace{4pt}}|} \hline
\mathsf{clg} \circ \mathsf{inst}^{\bullet}       & = & \mathsf{inst} \\
\mathsf{clg} \circ \mathsf{typ}^{\bullet}        & = & \mathsf{typ} \\
\mathsf{clsn} \circ \mathsf{inst}                & = & \mathsf{inst}^{\bullet} \\
\mathsf{clsn} \circ \mathsf{typ}                 & = & \mathsf{typ}^{\bullet} \\
\mathsf{id}_{\mathsf{CLSN}}                      & = & \mathsf{clg} \circ \mathsf{clsn} \\
\mathsf{id}_{\mathsf{CLG}}                       & \cong & \mathsf{clsn} \circ \mathsf{clg} \\
\iota^{\bullet} : \mathsf{inst}^{\bullet} & \Rightarrow & \mathsf{left}^{\bullet} \circ |{-}|^{\mathsf{op}} \\ 
\tau^{\bullet}  : \mathsf{typ}^{\bullet}  & \Rightarrow & \mathsf{right}^{\bullet} \circ |{-}| 
\\ \hline
\end{array}$
&
$\begin{array}[t]{|@{\hspace{4pt}}rcl@{\hspace{4pt}}|} \hline
|\mathsf{mod}|_{\Im} & = & \mathsf{clsn}_{\Im} \circ \mathsf{inst} \\
\mathsf{sen}_{\Im} & = & \mathsf{clsn}_{\Im} \circ \mathsf{typ} \\
\mathsf{clg}_{\Im} & = & \mathsf{clsn}_{\Im} \circ \mathsf{clg} \\
\mathsf{clsn}_{\Im} & = & \mathsf{clg}_{\Im} \circ \mathsf{clsn} \\
\mathsf{left}_{\Im}^{\bullet}  & = & \mathsf{clg}_{\Im} \circ \mathsf{left}^{\bullet} \\
\mathsf{right}_{\Im}^{\bullet}  & = & \mathsf{clg}_{\Im} \circ \mathsf{right}^{\bullet} \\
\iota_{\Im}^{\bullet} : |\mathsf{mod}|_{\Im} & \Rightarrow & \mathsf{left}_{\Im}^{\bullet} \circ |{-}|^{\mathsf{op}}
= \mathsf{clg}_{\Im} \circ \iota^{\bullet} \\ 
\tau_{\Im}^{\bullet} : \mathsf{sen}_{\Im} & \Rightarrow & \mathsf{right}_{\Im}^{\bullet} \circ |{-}|
= \mathsf{clg}_{\Im} \circ \tau^{\bullet} \\ \hline\hline
\mathsf{clo}_{\Im} : \mathsf{cpog}_{\Im} & \Rightarrow & \mathsf{clg}_{\Im} \circ \mathsf{incl} \\
\mathsf{left}_{\Im}^{\vdash}  & = & \mathsf{cpog}_{\Im} \circ \mathsf{left}^{\vdash} \\
\exists_{\Im} = \mathsf{right}_{\Im}^{\vdash}  & = & \mathsf{cpog}_{\Im} \circ \mathsf{right}^{\vdash} \\
\mathsf{Theory}_{\Im}^{\exists} & = & {\mathsf{Gr}(\exists_{\Im})}^{\mathsf{op}} \\ \hline
\end{array}$
\end{tabular}
\end{tabular}
\end{center}
\caption{The Architecture of Institutions}
\label{institutions-architecture}
\end{figure}
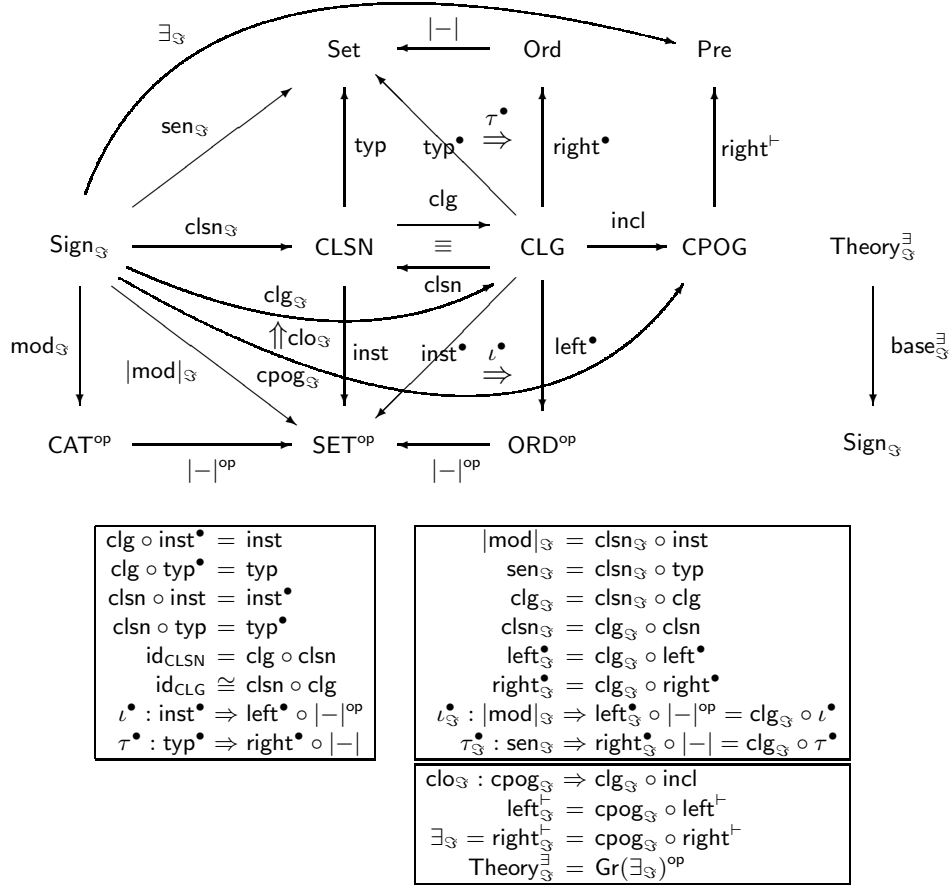

\subsection{Truth Construction}\label{subsec-truth-construction}

The components of an institution $\Im$,
which are illustrated on the left side of Figure~\ref{institutions-architecture},
can be packed together as a classification functor
$\mathsf{clsn}_{\Im} : \mathsf{Sign}_{\Im} \rightarrow \mathsf{CLSN}$
into the category of (large) classifications and infomorphisms \cite{barwise:seligman:97},
where for every signature $\Sigma$ 
the satisfaction relation forms the classification
$\mathsf{clsn}_{\Im}(\Sigma) = \langle |\mathsf{mod}|_{\Im}(\Sigma), \mathsf{sen}_{\Im}(\Sigma), \models_{\Im, \Sigma} \rangle$
and for every signature morphism $\sigma : \Sigma_{1} \rightarrow \Sigma_{2}$ 
the satisfaction condition states the fundamental condition for the infomorphism
$\mathsf{clsn}_{\Im}(\sigma) = \langle |\mathsf{mod}|_{\Im}(\sigma), \mathsf{sen}_{\Im}(\sigma) \rangle : \mathsf{clsn}_{\Im}(\Sigma_{1})\rightleftharpoons\mathsf{clsn}_{\Im}(\Sigma_{2})$.
However, the category of classifications $\mathsf{CLSN}$ is equivalent to the category of (large) concept lattices $\mathsf{CLG}$ (complete lattices with two-sided generators) and concept morphisms:
$\mathsf{CLSN} \equiv \mathsf{CLG}$ \cite{kent:02},
an equivalence mediated by 
the concept lattice functor $\mathsf{clg} : \mathsf{CLSN} \rightarrow \mathsf{CLG}$ and
the classification functor $\mathsf{clsn} : \mathsf{CLG} \rightarrow \mathsf{CLSN}$.
Several isomorphic representations of concept lattices are possible
--- a full two-sided version with both extents and intents, an extent-only version and an intent-only version.
In institution theory, we normally choose the intent-only version,
where $\mathsf{clg}_{\Im}(\Sigma)$ has the complete lattice component $\mathsf{th}_{\Im}^{\bullet}(\Sigma)$,
whose elements (formal concepts) are the closed $\Sigma$-theories of $\Im$.

Hence,
an alternate expression for an institution is 
a concept lattice functor $\mathsf{clg}_{\Im} : \mathsf{Sign}_{\Im} \rightarrow \mathsf{CLG}$,
where for every signature $\Sigma$ 
the associated concept lattice is the quintuple
$\mathsf{clg}_{\Im}(\Sigma)
= \langle |\mathsf{mod}|_{\Im}(\Sigma), \iota_{\Im}^{\bullet}(\Sigma), \mathsf{th}_{\Im}^{\bullet}(\Sigma), \tau_{\Im}^{\bullet}(\Sigma), \mathsf{sen}_{\Im}(\Sigma) \rangle$
and for every signature morphism $\sigma : \Sigma_{1} \rightarrow \Sigma_{2}$ 
the associated concept morphism is the quadruple
$\mathsf{clg}_{\Im}(\sigma)
= \langle |\mathsf{mod}|_{\Im}(\sigma), \mathsf{left}_{\Im}^{\bullet}(\sigma), \mathsf{right}_{\Im}^{\bullet}(\sigma), \mathsf{sen}_{\Im}(\sigma) \rangle : \mathsf{clg}_{\Im}(\Sigma_{1})\rightleftharpoons\mathsf{clg}_{\Im}(\Sigma_{2})$.
The left closed operator
$\mathsf{left}_{\Im}^{\bullet}(\sigma) : \mathsf{th}_{\Im}^{\bullet}(\Sigma_{1}) \leftarrow \mathsf{th}_{\Im}^{\bullet}(\Sigma_{2})$,
which is a join-preserving monotonic function,
is the substitution operator $\mathsf{subst}_{\Im}(\sigma) = {\mathsf{sen}_{\Im}(\sigma)}^{-1}$, restricted to closed theories.
The right closed operator
$\mathsf{right}_{\Im}^{\bullet}(\sigma) = \exists_{\Im}(\sigma) \cdot \mathsf{clo}_{\Im}(\Sigma_{2}) : \mathsf{th}_{\Im}^{\bullet}(\Sigma_{1}) \rightarrow \mathsf{th}_{\Im}^{\bullet}(\Sigma_{2})$,
which is a meet-preserving monotonic function,
is the existential quantification operator, restricted to closed theories, composed with the target closure operator.
The left-right pair of operators forms a Galois connection
$\phi_{\Im}^{\bullet}(\sigma)
= \langle \mathsf{left}_{\Im}^{\bullet}(\sigma), \mathsf{right}_{\Im}^{\bullet}(\sigma) \rangle : \mathsf{th}_{\Im}^{\bullet}(\Sigma_{2}) \rightleftharpoons \mathsf{th}_{\Im}^{\bullet}(\Sigma_{1})$.
The model operator is compatible with the left closed operator (via model-embedding):
$|\mathsf{mod}|_{\Im}(\sigma) \cdot \iota_{\Im}^{\bullet}(\Sigma_{1})
=
\iota_{\Im}^{\bullet}(\Sigma_{2}) \cdot \mathsf{left}_{\Im}^{\bullet}(\sigma)$,
and the sentence operator is compatible with the right closed operator (via sentence-embedding):
$\tau_{\Im}^{\bullet}(\Sigma_{1}) \cdot \mathsf{right}_{\Im}^{\bullet}(\sigma)
=
\mathsf{sen}_{\Im}(\sigma) \cdot \tau_{\Im}^{\bullet}(\Sigma_{2})$.
The equivalence of the two functors, $\mathsf{clsn}_{\Im}$ and $\mathsf{clg}_{\Im}$, is expressed by the identities:
$\mathsf{clg}_{\Im} = \mathsf{clsn}_{\Im} \circ \mathsf{clg}$
and
$\mathsf{clsn}_{\Im} = \mathsf{clg}_{\Im} \circ \mathsf{clsn}$.

\begin{sloppypar}
The category $\mathsf{CPOG}$ of (large) complete preorders with two-sided generators is adjointly related to the category $\mathsf{CLG}$ of concept lattices:
$\mathsf{CLG} \dashv\vdash \mathsf{CPOG}$,
an adjointness mediated by 
the inclusion functor $\mathsf{incl} : \mathsf{CLG} \rightarrow \mathsf{CPOG}$ and
the equivalence quotient functor $\mathsf{quo} : \mathsf{CPOG} \rightarrow \mathsf{CLG}$.
In $\mathsf{CPOG}$,
the orders do not necessarily satisfy antisymmetry --- a pair of elements may be equivalent but not equal.
An institution $\Im$ has an associated complete preorder functor
$\mathsf{cpog}_{\Im} : \mathsf{Sign}_{\Im} \rightarrow \mathsf{CPOG}$,
which is compositionally adjoint to the concept lattice functor (and hence, also to the classification functor).
For every signature $\Sigma$,
the associated complete preorder is the quintuple
$\mathsf{cpog}_{\Im}(\Sigma)
= \langle |\mathsf{mod}|_{\Im}(\Sigma), \mathsf{mod\mbox{-}th}_{\Im}(\Sigma), \mathsf{th}_{\Im}^{\vdash}(\Sigma), \mathsf{sen\mbox{-}th}_{\Im}(\Sigma), \mathsf{sen}_{\Im}(\Sigma) \rangle$
and for every signature morphism $\sigma : \Sigma_{1} \rightarrow \Sigma_{2}$ 
the associated complete preorder morphism is the quadruple
$\mathsf{cpog}_{\Im}(\sigma)
= \langle |\mathsf{mod}|_{\Im}(\sigma), \mathsf{left}_{\Im}^{\vdash}(\sigma), \mathsf{right}_{\Im}^{\vdash}(\sigma), \mathsf{sen}_{\Im}(\sigma) \rangle : \mathsf{cpog}_{\Im}(\Sigma_{1})\rightleftharpoons\mathsf{cpog}_{\Im}(\Sigma_{2})$.
For every signature morphism $\sigma : \Sigma_{1} \rightarrow \Sigma_{2}$,
the existential quantification monotonic function
$\exists_{\Im}(\sigma) : \mathsf{th}_{\Im}^{\vdash}(\Sigma_{1}) \rightarrow \mathsf{th}_{\Im}^{\vdash}(\Sigma_{2})$
maps a source theory $T_{1} \in \mathsf{th}_{\Im}(\Sigma_{1})$ to the target theory
$\exists_{\Im}(\sigma)(T_{1})
= \{ s_{2} \in \mathsf{sen}_{\Im}(\Sigma_{2}) \,|\, \exists_{s_{1} \in \mathsf{sen}_{\Im}(\Sigma_{1})}\, (\mathsf{sen}_{\Im}(\sigma)(s_{1}) = s_{2}) \;\&\; s_{1} \in T_{1} \} \in \mathsf{th}_{\Im}(\Sigma_{2})$,
and the substitution monotonic function is the inverse image of the sentence function
$\mathsf{subst}_{\Im}(\sigma) = {\mathsf{sen}_{\Im}(\sigma)}^{-1} : \mathsf{th}_{\Im}^{\vdash}(\Sigma_{1}) \leftarrow \mathsf{th}_{\Im}^{\vdash}(\Sigma_{2})$.
The left entailment operator
$\mathsf{left}_{\Im}^{\vdash}(\sigma)
= \mathsf{clo}_{\Im}(\Sigma_{2}) \cdot \mathsf{subst}_{\Im}(\sigma) : \mathsf{th}_{\Im}^{\vdash}(\Sigma_{1}) \leftarrow \mathsf{th}_{\Im}^{\vdash}(\Sigma_{2})$,
which is a join-preserving monotonic function,
is the composition of target closure with substitution.
The right entailment operator
$\mathsf{right}_{\Im}^{\vdash}(\sigma) = \exists_{\Im}(\sigma) : \mathsf{th}_{\Im}^{\vdash}(\Sigma_{1}) \rightarrow \mathsf{th}_{\Im}^{\vdash}(\Sigma_{2})$,
which is a meet-preserving monotonic function,
is existential quantification.
The left-right pair of operators forms a Galois connection
$\phi_{\Im}^{\vdash}(\sigma)
= \langle \mathsf{left}_{\Im}^{\vdash}(\sigma), \mathsf{right}_{\Im}^{\vdash}(\sigma) \rangle : \mathsf{th}_{\Im}^{\vdash}(\Sigma_{2}) \rightleftharpoons \mathsf{th}_{\Im}^{\vdash}(\Sigma_{1})$.
The model operator is compatible with the left entailment operator (via model-embedding):
$|\mathsf{mod}|_{\Im}(\sigma) \cdot \mathsf{mod\mbox{-}th}_{\Im}(\Sigma_{1})
= \mathsf{mod\mbox{-}th}_{\Im}(\Sigma_{2}) \cdot \mathsf{left}_{\Im}^{\vdash}(\sigma)$,
and the sentence operator is compatible with the right entailment operator (via sentence-embedding):
$\mathsf{sen\mbox{-}th}_{\Im}(\Sigma_{1}) \cdot \mathsf{right}_{\Im}^{\vdash}(\sigma)
= \mathsf{sen}_{\Im}(\sigma) \cdot \mathsf{sen\mbox{-}th}_{\Im}(\Sigma_{2})$.
\end{sloppypar}

For every signature $\Sigma$,
the closure operator 
$\mathsf{clo}_{\Im, \Sigma} : \mathsf{th}_{\Im}^{\vdash}(\Sigma) \rightarrow \mathsf{th}_{\Im}^{\bullet}(\Sigma)$
and inclusion
$\mathsf{incl}_{\Im, \Sigma} : \mathsf{th}_{\Im}^{\vdash}(\Sigma) \leftarrow \mathsf{th}_{\Im}^{\bullet}(\Sigma)$
are monotonic,
and form a Galois connection
$\rho_{\Im}(\Sigma)
= \langle \mathsf{incl}_{\Im}(\Sigma), \mathsf{clo}_{\Im}(\Sigma) \rangle : \mathsf{th}_{\Im}^{\bullet}(\Sigma) \rightleftharpoons \mathsf{th}_{\Im}^{\vdash}(\Sigma)$.
The left (entailment, closed) monotonic function commutes with the inclusion monotonic function:
$\mathsf{incl}_{\Im}(\Sigma_{2}) \cdot \mathsf{left}_{\Im}^{\vdash}(\sigma)
= \mathsf{left}_{\Im}^{\bullet}(\sigma) \cdot \mathsf{incl}_{\Im}(\Sigma_{1})$
for every signature morphism $\sigma : \Sigma_{1} \rightarrow \Sigma_{2}$.
The right (entailment, closed) monotonic function commutes with the closure monotonic function:
$\mathsf{right}_{\Im}^{\vdash}(\sigma) \cdot \mathsf{clo}_{\Im}(\Sigma_{2})
= \mathsf{clo}_{\Im}(\Sigma_{1}) \cdot \mathsf{right}_{\Im}^{\bullet}(\sigma)$
for every signature morphism $\sigma : \Sigma_{1} \rightarrow \Sigma_{2}$.
This means that the Galois connections commute
$\phi_{\Im}^{\bullet}(\sigma) \circ \rho_{\Im}(\Sigma_{1})
= \rho_{\Im}(\Sigma_{2}) \circ \phi_{\Im}^{\vdash}(\sigma)$
for every signature morphism $\sigma : \Sigma_{1} \rightarrow \Sigma_{2}$.
For every signature $\Sigma$,
closure (with a slight abuse of notation) forms a $\mathsf{CPOG}$-morphism
$\mathsf{clo}_{\Im}(\Sigma)
= \langle 1_{|\mathsf{mod}|_{\Im}(\Sigma)}, \mathsf{incl}_{\Im}(\Sigma), \mathsf{clo}_{\Im}(\Sigma), 1_{\mathsf{sen}_{\Im}(\Sigma)} \rangle : \mathsf{cpog}_{\Im}(\Sigma) \rightleftharpoons \mathsf{clg}_{\Im}(\Sigma)$.
For every signature morphism $\sigma : \Sigma_{1} \rightarrow \Sigma_{2}$,
we have the identity (commuting square)
$\mathsf{clo}_{\Im}(\Sigma_{1}) \circ \mathsf{clg}_{\Im}(\sigma)
= \mathsf{cpog}_{\Im}(\sigma) \circ \mathsf{clo}_{\Im}(\Sigma_{2})$.
Hence,
closure is a natural transformation
$\mathsf{clo}_{\Im} : \mathsf{cpog}_{\Im} \Rightarrow \mathsf{clg}_{\Im} \circ \mathsf{incl} : \mathsf{Sign}_{\Im} \rightarrow \mathsf{CPOG}$.

\subsection{Theories}

\begin{sloppypar}
An institution $\Im$ has an existential quantification theory functor (indexed category)
$\exists_{\Im} : (\mathsf{Sign}_{\Im}^{\mathsf{op}})^{\mathsf{op}} = \mathsf{Sign}_{\Im} \rightarrow \mathsf{Pre}$
to the category of (small) preorders and monotonic functions,
where
$\exists_{\Im}(\Sigma) = \mathsf{th}_{\Im}^{\vdash}(\Sigma)$
is the complete entailment preorder (``lattice of theories'') for every signature $\Sigma$,
and
$\exists_{\Im}(\sigma) : \mathsf{th}_{\Im}^{\vdash}(\Sigma_{1}) \rightarrow \mathsf{th}_{\Im}^{\vdash}(\Sigma_{2})$
is the direct existential image along the sentence function
for every signature morphism $\sigma : \Sigma_{1} \rightarrow \Sigma_{2}$.
The Grothendieck construction on this functor
produces a flattened category of theories
$\mathsf{Theory}_{\Im}^{\exists} = {\mathsf{Gr}(\exists_{\Im})}^{\mathsf{op}}$
(see the right side of Figure~\ref{institutions-architecture})
whose objects are pairs $\langle \Sigma, T \rangle$ where $T$ is a $\Sigma$-theory,
and whose morphisms
$\sigma : \langle \Sigma_{1}, T_{1} \rangle \rightarrow \langle \Sigma_{2}, T_{2} \rangle$
are signature morphsms
$\sigma : \Sigma_{1} \rightarrow \Sigma_{2}$
that map source axioms to target theorems
$T_{2}^{\bullet} \supseteq \exists_{\Im}(\sigma)(T_{1})$,
or equivalently satisfy the constraint
${\mathsf{sen}_{\Im}(\sigma)}^{-1}(T_{2}^{\bullet}) \supseteq T_{1}$. 
There is an obvious base signature projection functor
$\mathsf{base}_{\Im}^{\exists} : \mathsf{Theory}_{\Im}^{\exists} \rightarrow \mathsf{Sign}_{\Im}$.
Composition and identities for theory morphisms are defined in terms of their base signature morphisms.
Based upon the limit theorem of the third section of the paper \cite{tarlecki:burstall:goguen:91} on indexed categories and institutions,
if $\mathsf{Sign}_{\Im}$ the category of signatures is cocomplete,
then $\mathsf{Theory}_{\Im}^{\exists}$ the category of theories is cocomplete.
\end{sloppypar}

In the institutional approach of the IFF,
unpopulated ontologies (no instance data and no classifications) are represented by theories,
and the semantic integration of unpopulated ontologies is represented by the fusion (colimit construction) of theories.
Fusion in the category of theories is direct existential image flow followed by meet in the lattices of theories (base signature fibers).
The fusion construction is outlined in the following algorithm.
\begin{enumerate}
\item {\rmfamily\bfseries Alignment:} Informally, identify the theories to be used in the construction.
Decide on the semantic interconnection (semantic mapping) between theories.
This may involve the introduction of some additional mediating (reference) theories.
\begin{enumerate}
\item Formally, create a diagram of theories ${\cal T}$ of shape (indexing) graph $G$ that indicates this selection and interconnection. 
This diagram of theories is transient, since it will be used only for this computation. 
Other diagrams could be used for other fusion constructions.
\item Compute the base diagram of signatures ${\cal S} = \mathsf{base}_{\Im}^{\exists}({\cal T})$ with the same shape.
In more detail,
${\cal S}
= \mathsf{base}_{\Im}^{\exists}({\cal T})
= \{ {\cal S}_k \} + \{ {\cal S}_e : {\cal S}_m \rightarrow {\cal S}_n \}
= \{ \mathsf{base}_{\Im}^{\exists}({\cal T}_k) \} + \{ \mathsf{base}_{\Im}^{\exists}({\cal T}_e) : \mathsf{base}_{\Im}^{\exists}({\cal T}_m) \rightarrow \mathsf{base}_{\Im}^{\exists}({\cal T}_n) \}$.
Form the colimit (fusion) signature $\Sigma = \coprod {\cal S}$ of this diagram, 
with signature fusion cocone $\sigma : {\cal S} \Rightarrow \Sigma$.
In more detail, $\sigma = \{ \sigma_k : {\cal S}_k \rightarrow \Sigma \}$,
satisfying the conditions $\sigma_m = {\cal S}_e \cdot \sigma_n$ for $G$-edge $e : m \rightarrow n$.
Being the basis for theory colimits,
signature colimits are important.
They involve the two opposed processes of ``summing'' and ``quotienting''.
Summing can be characterized as ``keeping things apart'' and ``preserving distinctness'',
whereas quotienting  can be characterized as ``putting things together'', ``identification'' and ``synonymy''.
The ``things'' involved here are symbolic, and may involve relation type symbols, entity type symbols and the concepts that they denote.
\end{enumerate}
\item {\rmfamily\bfseries Unification:} Form the colimit (fusion) theory $T = \coprod {\cal T}$ of this diagram of theories,
with theory fusion cocone $\tau : {\cal T} \Rightarrow T$.
The fusion cocone is a universal corelation \cite{goguen:dagstuhl}
that connects the individual theories in the diagram to the fusion theory.
The fusion theory may be virtual.
\begin{enumerate}
\item Move (the individual theories $\{ {\cal T}_k \}$ in) the diagram of theories ${\cal T}$ 
from the ``lattice of theory diagrams'' $\mathsf{th}_{\Im}^{\vdash}({\cal S})$
along the signature morphisms in the signature fusion cocone $\sigma : {\cal S} \Rightarrow \Sigma$ 
to the lattice of theories $\mathsf{th}_{\Im}^{\vdash}(\Sigma)$ using the direct image function, 
getting the homogeneous diagram of theories $\exists_{\Im}(\sigma)({\cal T})$ with the same shape $G$, 
where each theory
$\exists_{\Im}(\sigma)({\cal T})_k = \exists_{\Im}(\sigma_k)({\cal T}_k)$
has the same base signature $\Sigma$ 
(the meaning of homogeneous).
\item Compute the meet (union) of the diagram $\exists_{\Im}(\sigma)({\cal T})$ within the ``lattice of theories'' $\mathsf{th}_{\Im}^{\vdash}(\Sigma)$,
getting the fusion theory
$T = \coprod {\cal T} 
= \bigwedge_{\Im, \Sigma} \exists_{\Im}(\sigma)({\cal T})
= \bigcup \exists_{\Im}(\sigma)({\cal T})$.
The signature fusion cocone is the base of the theory fusion cocone: 
$\sigma = \mathsf{base}_{\Im}(\tau) : \mathsf{base}_{\Im}({\cal T}) \Rightarrow \mathsf{base}_{\Im}(T)$.
\end{enumerate}
\end{enumerate}

\section{Institutions in the Information Flow Framework}

The main application of the IFF is institutional.
To be independent of the logic used,
the IFF represents and manipulates ontological structures within the metatheory of institutions.
The IFF has work-in-progress axiomatizations for the institutions and connecting institution morphisms
of information flow ({\bfseries IF}), equational logic ({\bfseries EQL}), order-sorted first order logic ({\bfseries FOL}) and the common logic standard ({\bfseries CL}), 
and is developing an axiomatization for the metatheory of institutions itself.
In particular,
the IFF has three work-in-progress efforts (meta-ontologies) that center around first order logic (FOL).
Two of these efforts ({\ttfamily IFF-ONT} and {\ttfamily IFF-OO}) are non-traditional and the third effort ({\ttfamily IFF-FOL}) is traditional.

\subsection{{\ttfamily IFF-ONT}}

The IFF Ontology (meta) Ontology ({\ttfamily IFF-ONT}) is an older non-traditional axiomatization for FOL,
It is based on the view that $n$-ary relations incorporate a notion of hypergraphs.
The {\ttfamily IFF-ONT} has finished axiomatizations for the concepts (and categories) of signature (aka language), theory, model and (local) logic.
The categories of signatures and theories are cocomplete.
Free models and logics exist.
{\ttfamily IFF-ONT} models are nonstandard:
one component is a subset of abstract tuples
(in the extreme, an {\ttfamily IFF-ONT} model might consist of just one abstract tuple),
and thus are much more flexible than ordinary FOL models;
and these models have a novel definition via classifications, $t \models r$ instead of $r(t)$.
However, a problem was discovered during the axiomatization of the {\ttfamily IFF-ONT}.
In order to represent an institution, model fiber (reduct) functors must be defined.
For these definitions, the variable component of the object functions of these model functors is required to be a bijection.
Hence, signatures and theories must be restricted to the corresponding subcategories.
However, these subcategories are not cocomplete.
But, completeness is a desirable property of the represented institution.
The problem was isolated to the requirement that variables in the {\ttfamily IFF-ONT} have a fixed sort.
This is apparently too inflexible, and in the newer effort, variables are sorted on the fly.

\subsection{{\ttfamily IFF-OO}}

The IFF Ontology (meta) Ontology ({\ttfamily IFF-OO}) is a newer non-traditional axiomatization for FOL.
It is still based on view that $n$-ary relations incorporate a notion of hypergraphs.
In fact, the {\ttfamily IFF-OO} identifies signatures (aka languages) with hypergraphs (of a certain kind).
This kind of hypergraph is flexible, since it does not use a reference (sort) function.
This means that node indices, which correspond to entity type variables, do not use a fixed sorting.
Model fiber (reduct) functors exist, and the categories of signatures and theories are cocomplete.
A special advantage has been found for this axiomatization: hypergraphs (signatures) are categorically equivalent to spangraphs.
Spangraphs, which are a more flexible definition of signature in terms of relational arity,
nicely model the common logic standard (CL) ``role-set syntax'' notation, where arguments form a set of role-value pairs. 
For example,
the sentence `{\ttfamily\small (Married (role-set: (wife Jill) (husband Jack)))}' expresses a marital relationship.
Pat Hayes (see the CL archive) has advocated this notation in CL development.
Position and argument order are not needed.
According to Hayes, this provides some insurance against communication errors.

\subsection{{\ttfamily IFF-FOL}}

\begin{figure}
\begin{center}
\setlength{\unitlength}{1pt}
\begin{picture}(200,150)(-20,-20)
\linethickness{1.0pt}
\put(66,66){\line(1,1){14}}
\put(16,16){\line(1,1){18}}
\put(86,16){\line(-1,1){18}}
\put(116,16){\line(1,1){14}}
\put(129,70){\line(-1,1){10.5}}
\put(100,30){\line(-1,-1){10}}
\put(100,30){\line(1,-1){10}}
\put(50,-20){\line(-1,-1){10}}
\put(50,-20){\line(1,-1){10}}
\put(70,81){\framebox(60,38){}}
\put(70,91){\makebox(60,38){\sffamily{FOL}}}
\put(70,81){\makebox(60,38){\sffamily{Language}}}
\put(70,71){\makebox(60,38){{\sffamily\scriptsize{(with equality)}}}}
\put(20,35){\framebox(60,30){}}
\put(20,40){\makebox(60,30){\sffamily{FOL}}}
\put(20,30){\makebox(60,30){\sffamily{Language}}}
\put(112,31){\framebox(76,38){}}
\put(112,41){\makebox(76,38){\sffamily{Equational}}}
\put(112,31){\makebox(76,38){\sffamily{Language}}}
\put(112,21){\makebox(76,38){{\sffamily\scriptsize{(universal algebra)}}}}
\put(-30,-15){\framebox(60,30){}}
\put(-30,-10){\makebox(60,30){\sffamily{Expression}}}
\put(-30,-20){\makebox(60,30){\sffamily{Language}}}
\put(70,-15){\framebox(60,30){}}
\put(70,-10){\makebox(60,30){\sffamily{Term}}}
\put(70,-20){\makebox(60,30){\sffamily{Language}}}
\end{picture}
\end{center}
\caption{The {\ttfamily IFF-FOL} module hierarchy}
\label{fol-hierarchy}
\end{figure}
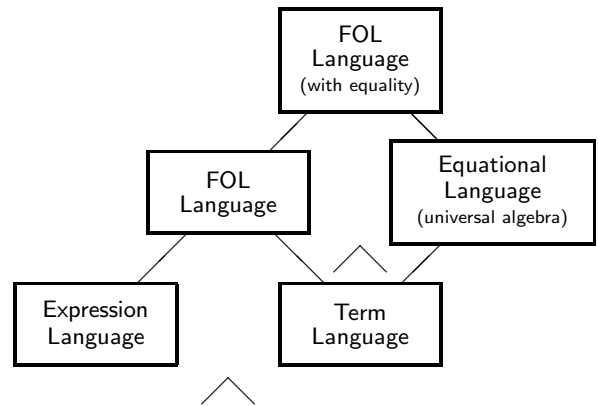

The IFF First Order Logic (meta) Ontology ({\ttfamily IFF-FOL}) gives a traditional axiomatization for FOL.
The {\ttfamily IFF-FOL} has a very modular architecture (Figure~\ref{fol-hierarchy}),
describing several institutions and institution morphisms along a spectrum,
including {\bfseries EQN} the institution for equational logic and {\bfseries FOL} the institution for first order logic. 
Each edge in the diagram of (Figure~\ref{fol-hierarchy}) is associated with institution morphisms in both directions,
projection downward and inclusion upward.
The central bifurcation is between terms and expressions.
FOL signatures are the pullback of expression signatures and term signatures over (bijections of) variables. 
FOL signatures with equality are the pullback of FOL signatures and universal algebra (equational signatures) over term signatures.
Term signatures consist of function (type) symbols and variables. 
The Lawvere construction is defined here. 
Equations can be added giving equational signatures (equational presentations) as an extension of term signatures. 
They define a quotient of their Lawvere category.
Expression signatures consist of relation (type) symbols and variables.
Peircian existential graphs can be included here.
FOL signatures consist of function (type) symbols, relation (type) symbols and variables.
From the modular perspective of Figure~\ref{fol-hierarchy},
an FOL signature is a term signature and an expression signature that share a common set of variables. 

\section{Summary}

The IFF is a descriptive category metatheory whose architecture contains nested metalanguages.
The main institutional application of the IFF is axiomatized in the Ontology (meta) Ontology ({\ttfamily IFF-ONT}, {\ttfamily IFF-OO}) and the FOL (meta) Ontology ({\ttfamily IFF-FOL}).
These form flexible and general institutions for first order logic.
Institutions formally express semantic integration as an ontological fusion process.
Each community represents their conceptual space in their own terms,
and connects with others via morphisms that enable ontological alignment specification.
Institutions represent logical environments and institution morphisms connect logical environments.
The IFF is in the process of axiomatizing the theory of institutions.


\end{document}